\documentclass[lettersize,journal]{IEEEtran}
\usepackage{amsmath,amsfonts}
\usepackage{algorithmic}
\usepackage{algorithm}
\usepackage{array}
\usepackage[caption=false,font=normalsize,labelfont=sf,textfont=sf]{subfig}
\usepackage{textcomp}
\usepackage{stfloats}
\usepackage{url}
\usepackage{verbatim}
\usepackage{graphicx}
\usepackage{cite}
\usepackage{lineno}

\begin{document}

\title{JiTTER: Jigsaw Temporal Transformer for Event Reconstruction \\ for Self-Supervised Sound Event Detection}

\author{Hyeonuk Nam,~\IEEEmembership{member,~IEEE,} Yong-Hwa Park,~\IEEEmembership{member,~IEEE,}
        % <-this % stops a space
\thanks{This paper was produced by the IEEE Publication Technology Group. They are in Piscataway, NJ.}% <-this % stops a space
\thanks{Manuscript received April 19, 2021; revised August 16, 2021.}}

% The paper headers
\markboth{Journal of \LaTeX\ Class Files,~Vol.~14, No.~8, August~2021}%
{Shell \MakeLowercase{\textit{et al.}}: A Sample Article Using IEEEtran.cls for IEEE Journals}

\IEEEpubid{0000--0000/00\$00.00~\copyright~2025 IEEE}
% Remember, if you use this you must call \IEEEpubidadjcol in the second
% column for its text to clear the IEEEpubid mark.

\maketitle

\begin{abstract}
Sound event detection (SED) has significantly benefited from self-supervised learning (SSL) approaches, particularly masked audio transformer for SED (MAT-SED), which leverages masked block prediction to reconstruct missing audio segments. However, while effective in capturing global dependencies, masked block prediction disrupts transient sound events and lacks explicit enforcement of temporal order, making it less suitable for fine-grained event boundary detection. To address these limitations, we propose \textit{JiTTER (Jigsaw Temporal Transformer for Event Reconstruction)}, an SSL framework designed to enhance temporal modeling in transformer-based SED. JiTTER introduces a hierarchical temporal shuffle reconstruction strategy, where audio sequences are randomly shuffled at both the block-level and frame-level, forcing the model to reconstruct the correct temporal order. This pretraining objective encourages the model to learn both global event structures and fine-grained transient details, improving its ability to detect events with sharp onset-offset characteristics. Additionally, we incorporate noise injection during block shuffle, providing a subtle perturbation mechanism that further regularizes feature learning and enhances model robustness. Experimental results on the DESED dataset demonstrate that JiTTER outperforms MAT-SED, achieving a 5.89\% improvement in PSDS, highlighting the effectiveness of explicit temporal reasoning in SSL-based SED. Our findings suggest that structured temporal reconstruction tasks, rather than simple masked prediction, offer a more effective pretraining paradigm for sound event representation learning. 
\end{abstract}


\section{Introduction}
\label{sec:intro}
Sound event detection (SED) is a fundamental task in machine listening and plays a crucial role in auditory intelligence, enabling applications in AI-driven perception, smart environments, and bioacoustic monitoring \cite{CASSE, DCASEtask4, crnn, sedmetrics, PSDS, freqdepinternoise}. In addition to SED, various research efforts have focused on speech and speaker recognition \cite{specaug, conformer, mpc, wav2vec2.0, hubert, ASP, SAP, tdyaccess, freqse, c2datt}, sound event recognition\cite{PANN,coughcam,etri,AST,beats},  sound event localization and detection (SELD) \cite{seld2019,starss22,2022t3report}, automated audio captioning (AAC) \cite{dcaseaac,clotho,chatgptaugaac}, few-shot bioacoustic detection \cite{dcasebed2024,bioacousticstrf}, human auditory perception \cite{prtfnet, brainstem}, highlighting the broad impact of auditory perception in real-world applications. Furthermore, recent advancements in sound synthesis \cite{audioldm,audiogen,vifs} have explored the generative modeling of sound events from textual descriptions, providing new perspectives in sound representation learning.

\begin{figure}[t]
   \centering
   \includegraphics[width=0.9\linewidth]{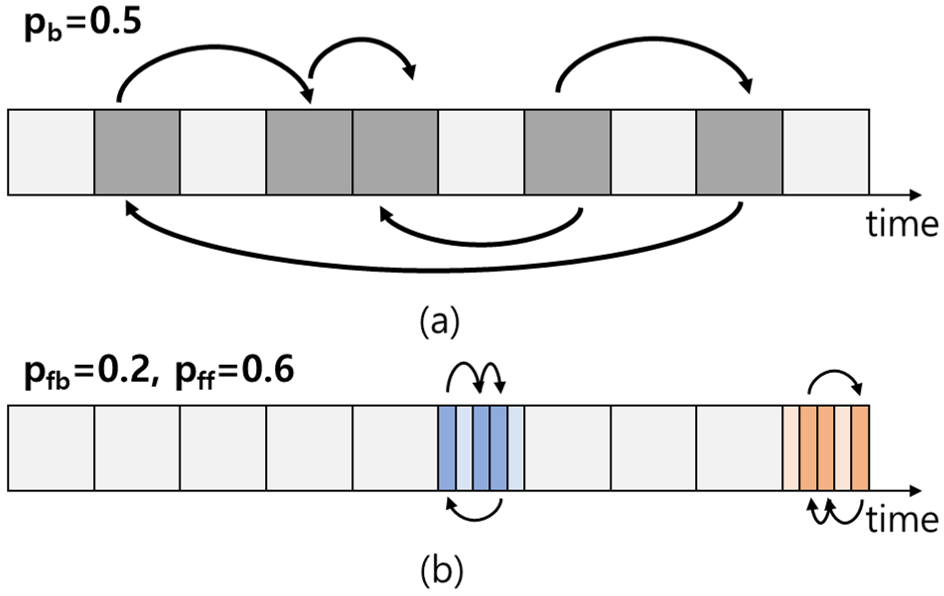}
   \caption{
   Illustration of the hierarchical temporal shuffle strategy in JiTTER, designed to improve temporal modeling for self-supervised SED.  
   (a) Block-Level Shuffle: The input audio sequence is divided into non-overlapping blocks, and a portion of these blocks (in darker grey) is randomly shuffled along the time axis. This disrupts global event dependencies while preserving intra-block structures, forcing the model to reconstruct event sequences at a higher level.  
   (b) Frame-Level Shuffle: A subset of blocks is randomly selected, and within each selected block (in blue and orange), a fraction of frames (in darker blue and orange) is randomly shuffled. This introduces fine-grained perturbations while maintaining the overall event order, helping the model learn transient sound characteristics.  
   Together, these two levels of shuffle perturbations encourage the model to reconstruct the correct temporal order, improving both global event structure comprehension and fine-grained boundary detection in SED.}
   \label{fig:jitter}
\end{figure}

SED aims to classify sound events while precisely localizing their temporal boundaries, making it a fundamental research area in auditory intelligence. Recent advances in deep learning have significantly improved SED performance \cite{dcase2020_1st, dcase2021_1st, mytechreport, dcase2023a_1st, dcase2024_1st, atstsed, matsed, PMAM}, driven by both architectural innovations and insights from auditory cognition \cite{filtaug,FDY,freqdeptalsp,stftchallenge, stftworkshop}. Early approaches primarily relied on convolutional neural networks (CNNs) to model spectral and temporal patterns in audio signals \cite{freqatt, DFD, PFD}. While CNN-based models demonstrated strong feature extraction capabilities, they struggled to capture long-range dependencies, which are essential for distinguishing between sequential and overlapping sound events. To overcome this limitation, transformer-based architectures pretrained on large-scale datasets such as AudioSet have been explored, enabling more effective temporal modeling \cite{audioset, dcase2023a_1st, dcase2024_1st, atstsed, dcase2024myworkshop}. These transformer-based models leverage self-attention mechanisms to capture global contextual information, leading to significant improvements in SED accuracy and robustness.

\IEEEpubidadjcol

Among recent transformer-based approaches, the masked audio transformer for SED (MAT-SED) introduced a self-supervised learning (SSL) strategy on full transformer model known as masked block prediction \cite{matsed}. This method involves masking temporal blocks of audio input and training the model to reconstruct the missing features, improving its ability to learn meaningful audio representations. MAT-SED employs the patchout fast spectrogram transformer (PaSST) as an encoder network and a transformer with relative positional encoding (RPE) as a context network, forming a fully transformer-based SED model \cite{passt}. The prototype-based masked audio model (PMAM) further extended this approach by introducing Gaussian mixture model (GMM)-based prototypical representations as semantically rich frame-level pseudo labels \cite{PMAM}. While these methods enhance the model's ability to capture long-range dependencies and improve generalization, they exhibit two fundamental limitations: transient event loss and lacking explicit temporal order constraints.

To overcome these challenges, we propose \textit{JiTTER (Jigsaw Temporal Transformer for Event Reconstruction)}, a self-supervised learning framework that introduces hierarchical temporal shuffle reconstruction to improve temporal modeling in SED. Unlike masked block prediction, which removes information, JiTTER shuffles audio segments at different temporal scales, forcing the model to learn to reconstruct their correct order. As shown in Figure \ref{fig:jitter}, JiTTER applies two levels of perturbation:
\begin{itemize}
    \item Block-Level Shuffle: Randomly shuffles non-overlapping blocks of audio, disrupting global event structures while preserving local coherence.
    \item Frame-Level Shuffle: Randomly selects blocks and shuffles a fraction of frames within each selected block, perturbing fine-grained temporal information.
\end{itemize}
By solving this hierarchical jigsaw-like reconstruction task, JiTTER explicitly enforces global event continuity and transient event representation, enhancing the model's ability to recognize event boundaries while capturing long-range dependencies. Additionally, we introduce noise injection during block shuffle to provide a controlled level of information corruption. Unlike masked block prediction, which completely removes event cues, noise injection partially obscures information while retaining weak structural signals, promoting more robust feature learning.

Our experimental results on the Domestic Environment Sound Event Detection (DESED) dataset demonstrate that JiTTER achieves a 5.89\% improvement in PSDS over MAT-SED, highlighting the importance of explicitly modeling temporal dependencies in self-supervised pretraining for SED. Moreover, ablation studies confirm the optimal shuffle configurations, suggesting that block-level shuffle primarily captures long-range temporal dependencies, while frame-level shuffle enhances short-range temporal structures. Multitask learning, which combines both shuffle strategies, yields the highest performance gains, further indicating that these perturbations operate at different temporal scales. This highlights the benefit of integrating multiple levels of temporal reordering to strengthen event representations.

The key contributions of this paper are:
\begin{enumerate}
    \item We introduce JiTTER, a self-supervised pretraining framework that enhances SED performance by reconstructing temporally shuffled sequences.
    \item We propose hierarchical temporal shuffle reconstruction, combining block-level and frame-level shuffling to improve global event continuity and transient event recognition.
    \item We incorporate noise injection during block shuffle to enhance feature learning while preserving weak structural cues.
    \item Extensive ablation studies validate the effectiveness of different shuffle configurations and multitask learning in improving SED performance.
    \item Our experiments on the DESED dataset demonstrate that JiTTER outperforms conventional masked prediction methods, achieving a 5.89\% PSDS improvement over MAT-SED.
\end{enumerate}
The official implementation code is available on GitHub\footnote{https://github.com/frednam93/JiTTER-SED}.

\section{Related Works}
\label{sec:relatedworks}
\subsection{Self-Supervised Learning for Sound Recognition}

SSL has been widely explored for audio tagging and general-purpose audio representation learning. SSL enables models to learn meaningful features without requiring explicit human annotations, making it particularly valuable in domains with limited labeled data. Several SSL approaches have been proposed to enhance audio tagging, focusing on capturing high-level acoustic representations.

BYOL-A \cite{byola} applies contrastive learning to augmented audio segments, ensuring consistency across different transformations. SSAST, an extension of the audio spectrogram transformer (AST), combines discriminative and generative masked spectrogram patch modeling to improve generalization across diverse audio tasks \cite{AST, ssast}. AudioMAE employs a masked autoencoder framework to reconstruct spectrogram patches using a transformer encoder-decoder \cite{audioMAE}, while BEATs iteratively optimizes an acoustic tokenizer alongside an SSL model to refine bidirectional audio representations \cite{beats}.

While these SSL methods have shown success in general-purpose audio learning and audio tagging, they are not specifically designed for SED, which requires precise event boundary localization. Unlike audio tagging, where the objective is to classify the presence of sound categories, SED demands accurate onset-offset detection and a structured understanding of sequential dependencies. Many existing SSL models prioritize general acoustic feature extraction but lack mechanisms to capture fine-grained temporal structures necessary for event-level modeling.

To address these challenges, SSL techniques specifically designed for SED have been proposed. MAT-SED introduced masked block prediction, where randomly selected temporal blocks in an audio sequence are masked, and the model is trained to reconstruct the missing features \cite{matsed}. This approach enhances the model’s ability to capture global contextual dependencies and improves its robustness to missing information. The prototype-based masked audio model (PMAM) further extended this technique by incorporating Gaussian mixture model (GMM)-based pseudo labels, providing additional semantic guidance during reconstruction \cite{PMAM}. These methods have demonstrated state-of-the-art performance in modeling long-range dependencies for self-supervised SED.

\subsection{Shuffle-Based Learning in Representation Learning}

Segment reordering tasks have been widely explored in self-supervised representation learning, particularly in vision and video processing. Jigsaw pretext tasks involve permuting image patches and training models to recover the correct order, leading to stronger spatial feature learning \cite{jigsawvision, jigswavit}. In the video domain, Shuffle and Learn  demonstrated that training models to predict the correct sequence of shuffled frames enhances motion feature learning \cite{shuffleandlearn}.

In audio processing, shuffle-based learning has been largely unexplored, with existing self-supervised approaches like TERA and BYOL-A focusing on temporal coherence rather than explicit sequence reconstruction. TERA promotes temporal consistency by reconstructing spectrograms altered along time, frequency, and magnitude axes, ensuring robustness to distortions but without enforcing strict sequential dependencies \cite{tera}. BYOL-A enhances temporal coherence through random resized cropping (RRC), which approximates pitch shifting and time stretching by randomly resizing and cropping spectrogram segments. This forces the model to learn representations that remain stable across variations in time and frequency \cite{byola}. However, neither method explicitly enforces event ordering, which is critical for SED. In contrast, JiTTER extends shuffle-based learning to SED by introducing hierarchical temporal shuffle reconstruction, where models must recover the correct sequence after structured perturbations at both block and frame levels. Unlike contrastive learning, JiTTER does not rely on negative pairs but directly optimizes for sequence restoration, making it more suitable for fine-grained temporal reasoning in SED.

\subsection{Temporal Modeling for Sound Event Detection}

Temporal modeling is critical in SED, where accurately capturing event onsets and offsets is essential. Transformer-based architectures have demonstrated strong capabilities in modeling long-range dependencies. Patchout fast spectrogram transformer (PaSST) \cite{passt} improves computational efficiency while preserving contextual modeling, making it a strong backbone for SED. More recently, studies have explored relative positional encoding (RPE) \cite{rpe}, which helps capture fine-grained temporal relationships.

A key challenge in SED is distinguishing overlapping events from sequentially occurring ones. Previous works have attempted to improve event boundary detection through classification-based segmentation \cite{YOHO} and data augmentation techniques \cite{specaug,mytechreport}, but these approaches lack a structured learning objective for temporal order recovery. JiTTER addresses this gap by enforcing explicit temporal reasoning through hierarchical shuffle-based reconstruction, leading to improved event boundary detection and reduced reliance on post-processing techniques.

\section{Methodology}
\label{sec:method}
\textit{JiTTER (Jigsaw Temporal Transformer for Event Reconstruction)} is a self-supervised learning framework designed to improve temporal representation learning for SED. Unlike masked prediction, JiTTER introduces a hierarchical temporal shuffle reconstruction strategy, challenging the model to recover the correct sequence of shuffled segments. This enhances the model’s ability to capture both local transient structures and long-term event dependencies, leading to improved sound event boundary detection.

\subsection{Limitations of Masked Block Prediction}

Masked block prediction, as employed in MAT-SED, removes entire audio segments, leading to the loss of transient sound events such as footsteps, door slams, and alarms, which are brief and temporally localized \cite{matsed}. Since these masked segments are absent during training, the model is forced to reconstruct them solely from surrounding context. This often results in inaccurate reconstructions, as the model may over-rely on background noise or unrelated acoustic cues instead of learning to capture fine-grained transient event structures. The absence of direct supervision on these short-duration events reduces the model’s ability to accurately detect event onsets and offsets, which is crucial for SED.

Additionally, masked block prediction does not explicitly enforce temporal order learning, making it less effective in distinguishing between overlapping and sequential events. Transformers inherently model attention-based relationships, but without explicit temporal constraints, they may struggle to recover the correct event sequence when multiple events occur in succession. This limitation weakens the model’s ability to differentiate between events that share similar spectral characteristics but differ in their temporal positioning. As a result, event boundaries may become blurred, reducing the precision of SED predictions.

To address these issues, JiTTER introduces hierarchical temporal shuffle reconstruction, which perturbs audio sequences at both the block and frame levels while preserving all content. Unlike masked block prediction, which removes information, JiTTER retains all temporal data but disrupts its order, forcing the model to reconstruct the correct sequence. This structured perturbation explicitly enforces temporal coherence, improving two critical aspects of SED: event boundary precision and transient event recognition.

By preserving all acoustic information while shuffling its order, JiTTER eliminates the risk of interpolation artifacts and ensures that the model learns event-level temporal dependencies rather than relying on surrounding context for reconstruction. This approach enhances both global event continuity and fine-grained transient modeling, leading to more accurate SED performance compared to masked block prediction.

\subsection{Hierarchical Temporal Shuffle Reconstruction}

JiTTER consists of two levels of temporal perturbation: block-level shuffle and frame-level shuffle, which are applied in parallel within a single forward pass. These perturbations operate at different temporal scales: block-level shuffle disrupts global event structures, while frame-level shuffle introduces local variations. The objective is to learn robust representations by solving a temporal jigsaw puzzle that requires the model to infer both long-range and fine-grained event dependencies.

\subsubsection{Block-Level Shuffle}
Block-level shuffle (Figure \ref{fig:jitter}a) partitions an audio sequence into non-overlapping blocks and randomly shuffles a portion (\( p_b \)), keeping the first and last blocks fixed as anchors.

Given an input audio sequence \( X \) of length \( T \), it is divided into \( B \) blocks:
\begin{equation}
    X = \{B_1, B_2, ..., B_B\}
\end{equation}
where each block \( B_i \) consists of multiple consecutive frames. One example of a randomly shuffled sequence is:
\begin{equation}
    \tilde{X}_b = \{B_1, B_2, B_{B-3}, B_4, ..., B_3, B_B\}
\end{equation}
where \( \tilde{X}_b \) represents a sequence perturbed by block-level shuffling. The model is pretrained to recover the correct order \( X \) from \( \tilde{X}_b \), which encourages the transformer to capture global contextual dependencies.

Inspired by masked prediction, we introduce Gaussian noise injection into shuffled blocks, partially obscuring information without full masking \cite{matsed, PMAM}. The goal is to obscure information slightly rather than removing it completely, allowing the model to retain weak structural cues while still reinforcing its ability to restore the original sequence.

Given a shuffled block \( B_i \), noise injection is performed as:
\begin{equation}
    B_i^{\text{noise}} = B_i + \lambda N_i, \quad N_i \sim \mathcal{N}(0, I)
\end{equation}
where \( N_i \) is Gaussian noise sampled from a normal distribution with zero mean and unit variance. Here, \( N_i \) has the same dimensions as \( B_i \), ensuring that noise is applied consistently across the entire block. The scaling factor \( \lambda \) controls the intensity of perturbation, and we set \( \lambda = 0.1 \) to introduce subtle distortion while preserving core temporal structures. This additional perturbation encourages the model to be more robust to small variations in real-world recordings while still maintaining the integrity of the shuffled reconstruction task.

\subsubsection{Frame-Level Shuffle}
As shown in Figure \ref{fig:jitter} (b), frame-level shuffling randomly selects a subset of blocks and applies intra-block frame shuffling. The hyperparameters \( p_{fb} \) and \( p_{ff} \) denote the proportion of blocks undergoing frame shuffling and the proportion of frames shuffled within each selected block, respectively.

For a chosen block \( B_i = \{f_{i1}, f_{i2}, ..., f_{iF}\} \), a portion of its frames is randomly permuted as follows:
\begin{equation}
    \tilde{B}_i = \{f_{i3}, f_{i2}, f_{iF}, f_{i4}, ..., f_{i1}\}
\end{equation}
where \( \tilde{B}_i \) represents an example of a shuffled block. The perturbed sequence \( \tilde{X}_f \) is then reconstructed by combining all modified blocks:
\begin{equation}
    \tilde{X}_f = \{\tilde{B}_1, B_2, ..., \tilde{B}_m, B_{m+1}, ..., B_B\}
\end{equation}
where \( m \) represents the number of shuffled blocks and \( \tilde{X}_f \) represents an example of a frame-level shuffled sequence. Unlike block shuffle, this preserves global order while introducing local perturbations, enhancing fine-grained temporal modeling.

\subsection{Training Objective}

JiTTER is trained using a reconstruction loss that explicitly enforces temporal structure learning by requiring the model to restore the correct event sequence from perturbed versions \( \tilde{X}_b \) and \( \tilde{X}_f \). Unlike masked block prediction, which removes information entirely, JiTTER preserves all temporal data but disrupts their order, making reconstruction a more structured learning objective.

The reconstruction loss is formulated as follows:
\begin{equation}
    \mathcal{L}_{\text{rec}}(\tilde{X}, X) = \sum_{t=1}^{T} || \mathcal{F}_{\theta}(\tilde{X})(t) - X(t) ||^2
\end{equation}
where \( X(t) \) represents the \( t \)-th frame of the original sequence, and \( \mathcal{F}_{\theta} \) is the transformer network parameterized by \( \theta \), which aims to reconstruct \( X \) from its shuffled counterpart \( \tilde{X} \). The model is optimized to recover both global event continuity and local transient details, leading to stronger representations for SED.

JiTTER's overall training objective is defined as:
\begin{equation}
    \mathcal{L}_{\text{JiTTER}} = \mathcal{L}_{\text{rec}}(\tilde{X}_b, X) + \mathcal{L}_{\text{rec}}(\tilde{X}_f, X)
\end{equation}
where \( \tilde{X}_b \) and \( \tilde{X}_f \) correspond to block-shuffled and frame-shuffled sequences, respectively.

This structured loss function ensures that:
\begin{itemize}
    \item Block-level shuffle encourages learning long-range temporal dependencies by forcing the model to recover the correct high-level sequence order.
    \item Frame-level shuffle improves fine-grained event localization by requiring the model to reconstruct disrupted transient event patterns.
\end{itemize}

Multitask learning, which combines both perturbations, allows the model to integrate global and local event structures effectively, leading to more robust temporal reasoning in SED.

\section{Experimental Setups}
\label{sec:setup}
This section describes the dataset, input feature extraction process, model architecture, training procedure, and evaluation metrics used in our experiments.

\subsection{Dataset}

We train, validate, and evaluate our models using the Domestic Environment Sound Event Detection (DESED) dataset \cite{DCASEtask4}, a widely used benchmark for domestic sound event detection. The dataset consists of 10-second-long audio clips recorded at a sampling rate of 16 kHz. It includes both synthetic and real recordings that simulate common household acoustic environments, covering ten sound event classes such as alarms, speech, running water, and vacuum cleaners.

The DESED dataset is divided into four subsets:
\begin{itemize}
    \item Strongly labeled synthetic data: Includes precise onset and offset annotations for each event.
    \item Weakly labeled real data: Indicates only the presence of events in each clip without specifying time boundaries.
    \item Unlabeled real data: Used in a self-supervised and semi-supervised learning setup to improve generalization.
    \item Strongly labeled real validation data: Reserved for model evaluation.
\end{itemize}
To ensure a fair comparison, we follow the standard Detection and Classification of Acoustic Scenes and Events (DCASE) 2021 Task 4 data partitioning scheme \cite{DCASEtask4}. The strongly labeled synthetic data is used for supervised learning, while weakly labeled and unlabeled data are leveraged in semi-supervised training. We also include the real strongly labeled training data from the DCASE 2022 Task 4 Challenge.

\subsection{Input Feature Extraction}

Raw audio waveforms are first normalized to a maximum absolute value of one to standardize input levels. We then extract log-mel spectrograms as input features using a short-time Fourier transform (STFT) with FFT size of 1024, Hop length of 320 (corresponding to 20 ms at 16 kHz), Hanning windowing function and 128 mel frequency bins. The resulting log-mel spectrograms serve as input to JiTTER’s transformer-based architecture.

\subsection{Data Augmentation}

To enhance the model's robustness against environmental variability, we apply a diverse set of augmentation techniques for fine-tuning stages:
\begin{itemize}
    \item Frame shift \cite{DCASEtask4}: Shifts the audio features by a small number of frames to introduce temporal variations.
    \item Mixup \cite{mixup}: Linearly interpolates pairs of spectrograms and labels, encouraging smoother decision boundaries.
    \item Time masking \cite{specaug}: Randomly masks sections of the time axis, simulating missing event cues.
    \item FilterAugment \cite{filtaug}: Randomly reweights frequency regions to simulate different acoustic environments.
    \item Frequency Distortion \cite{matsed}: Perturbs frequency components to improve spectral robustness.
\end{itemize}
We apply Mixup to both strongly and weakly labeled datasets, while time masking is applied jointly to the input spectrogram and its corresponding labels to maintain consistency.

\subsection{Model Architecture}

JiTTER extends the MAT-SED framework by replacing masked block prediction with hierarchical temporal shuffle reconstruction. The model consists of:
\begin{itemize}
    \item A patchout fast spectrogram transformer (PaSST) encoder, which extracts rich spectral-temporal representations \cite{passt}.
    \item A Transformer-based context network with relative positional encoding (RPE), designed to capture long-range temporal dependencies \cite{rpe}.
    \item A fully connected (FC) classification head, which predicts frame-wise event occurrences.
\end{itemize}
Unlike conventional masked prediction, JiTTER applies block-level and frame-level shuffle perturbations before input sequences enter the transformer model. The network is then pretrained to reconstruct the original sequence, enforcing temporal structure learning.

\subsection{Training Procedure}

JiTTER follows a three-stage training paradigm to progressively refine temporal representations for SED:
\begin{enumerate}
    \item \textbf{Pretraining}: The context network, consisting of the transformer with relative positional encoding (RPE), is trained using the hierarchical temporal shuffle strategy to reconstruct shuffled sequences. During this phase, the context network is updated, while the encoder network (PaSST) remains frozen. This stage runs for 6000 steps.
    \item \textbf{Feature adaptation}: The pretrained transformer remains fixed, while the SED and AT heads are trained separately using the SED objective. This step allows the classification layers to adapt to structured temporal representations. Training runs for an additional 6000 steps.
    \item \textbf{Fine-tuning}: The entire model, including the transformer, PaSST encoder, SED, and AT heads, is jointly optimized in an end-to-end manner with the SED objective to refine overall event detection performance. This final stage runs for another 6000 steps.
\end{enumerate}
All training is conducted on NVIDIA RTX 3090 GPUs using the AdamW optimizer with a weight decay of \(1 \times 10^{-4}\).

\subsection{Loss Function}

JiTTER is trained using a multi-stage loss function that combines self-supervised reconstruction loss for pretraining and semi-supervised classification loss for SED fine-tuning.

\subsubsection{Pretraining Loss}
During the pretraining stage, JiTTER is optimized with a reconstruction loss that encourages the model to infer the original sequence from temporally shuffled versions. The objective is formulated as:
\begin{equation}
    \mathcal{L}_{\text{rec}}(\tilde{X}, X) = \sum_{t=1}^{T} || \mathcal{F}_{\theta}(\tilde{X})(t) - X(t) ||^2
\end{equation}
where \( X(t) \) is the \( t \)-th frame of the original sequence, and \( \mathcal{F}_{\theta}(\tilde{X})(t) \) is the model’s reconstructed prediction. JiTTER optimizes both block-shuffled and frame-shuffled sequences:
\begin{equation}
    \mathcal{L}_{\text{JiTTER}} = \mathcal{L}_{\text{rec}}(\tilde{X}_b, X) + \mathcal{L}_{\text{rec}}(\tilde{X}_f, X)
\end{equation}
where \( \tilde{X}_b \) and \( \tilde{X}_f \) denote the block-level and frame-level shuffled sequences, respectively.

\subsubsection{SED Fine-Tuning Loss}
After pretraining, JiTTER is optimized using a supervised classification loss for polyphonic SED. Following standard SED frameworks \cite{matsed, PMAM}, the loss function consists of:
\begin{itemize}
    \item Strong classification loss (\( L_s \)): Applied to strongly labeled data using binary cross-entropy (BCE).
    \item Weak classification loss (\( L_w \)): Applied to weakly labeled data using BCE.
    \item Consistency loss (\( L_c \)): Ensures consistency between the student and teacher models using mean square error (MSE).
\end{itemize}

The overall fine-tuning loss is defined as:
\begin{equation}
    L_{\text{SED}} = B(S_P, l_s) + w_W B(W_P, l_w) + w_C L_{\text{cons}}
\end{equation}
where \( S_P, W_P \) are the strong and weak predictions from the student model, \( l_s, l_w \) are the corresponding ground-truth labels, and \( w_W \) and \( w_C \) are weighting factors for weak classification and consistency losses.

The consistency loss \( L_{\text{cons}} \) is given by:
\begin{equation}
    L_{\text{cons}} = M(S_P, sg(S_P^T)) + M(W_P, sg(W_P^T))
\end{equation}
where \( sg(\cdot) \) denotes a stop-gradient operation, and \( S_P^T, W_P^T \) are the strong and weak predictions from the teacher model.

This multi-stage optimization strategy ensures that JiTTER first learns robust temporal representations via self-supervised pretraining, which are later refined with supervised event classification.

\subsection{Post-Processing}

To refine event predictions, we apply a post-processing pipeline consisting of two steps. First, weak prediction masking is applied, where strong predictions are retained only if their confidence surpasses the corresponding weak predictions \cite{mytechreport}. This ensures that detected events are supported by global event presence information. Next, median filtering is used to smooth predictions and reduce spurious detections. Specifically, we apply median filters with a window size of 5 for transient sound events and a window size of 20 for stationary sound events. This strategy helps to suppress short-duration false positives for transient events while preventing abrupt fluctuations in stationary event predictions.

\subsection{Evaluation Metrics}

To evaluate SED performance, the polyphonic sound detection score (PSDS) was used \cite{PSDS}. PSDS is a metric specifically designed for evaluating polyphonic SED systems, addressing the limitations of conventional collar-based event F-scores and event error rates by incorporating intersection-based event detection. In addition, unlike traditional metrics, PSDS considers the full polyphonic receiver operating characteristic (ROC) curve, summarizing system performance across multiple operating points. This allows for a more robust and comprehensive assessment of an SED system’s capabilities, making it less sensitive to subjective annotation errors and more suitable for real-world applications. Additionally, PSDS provides better insights into classification stability across different sound classes and dataset biases.

For the DCASE Challenge 2021, 2022, and 2023 Task 4 benchmarks \cite{DCASEtask4}, two types of PSDS were employed: PSDS1 and PSDS2. Among them, PSDS2 is more aligned with the audio tagging task rather than SED, as it emphasizes the classification of event presence rather than precise temporal localization \cite{mytechreport, SEBB}. Therefore, we report only PSDS1 in this work, as it directly measures an SED system's ability to detect sound events with accurate onset and offset timings. PSDS values reported in the tables represent the best scores obtained from six independent training runs, ensuring that the evaluation reflects a reliable and well-optimized performance estimate of the proposed model.

\section{Results and Discussion}
\label{sec:result}
\begin{table}[t]
    \centering
    \caption{Ablation study on JiTTER using block-level shuffle, frame-level shuffle, and multitask learning. PSDS values are averaged over six independent training runs.}
    \label{tab:ablation}
    \begin{tabular}{l | c c c | c}
        \hline
        \textbf{Method} & \( p_b \) & \( p_{fb} \) & \( p_{ff} \) & \textbf{PSDS} (\(\uparrow\)) \\
        \hline
        MAT-SED (Baseline)  & -    & -    & -    & 0.543 \\
        \hline
        \multicolumn{5}{c}{\textbf{Block-Level Shuffle}} \\
        \hline
        JiTTER (Block Shuffle) & 0.25 & -    & -    & 0.566 \\
        JiTTER (Block Shuffle) & 0.5  & -    & -    & 0.569 \\
        JiTTER (Block Shuffle) & 0.75 & -    & -    & \textbf{0.570} \\
        \hline
        \multicolumn{5}{c}{\textbf{Frame-Level Shuffle}} \\
        \hline
        JiTTER (Frame Shuffle) & -    & 0.25 & 0.25 & 0.560 \\
        JiTTER (Frame Shuffle) & -    & 0.25 & 0.5  & 0.563 \\
        JiTTER (Frame Shuffle) & -    & 0.25 & 0.75 & 0.563 \\
        JiTTER (Frame Shuffle) & -    & 0.5  & 0.25 & \textbf{0.564} \\
        JiTTER (Frame Shuffle) & -    & 0.5  & 0.5  & 0.562 \\
        JiTTER (Frame Shuffle) & -    & 0.5  & 0.75 & 0.563 \\
        JiTTER (Frame Shuffle) & -    & 0.75 & 0.25 & 0.563 \\
        JiTTER (Frame Shuffle) & -    & 0.75 & 0.5  & 0.562 \\
        JiTTER (Frame Shuffle) & -    & 0.75 & 0.75 & 0.562 \\
        \hline
        \multicolumn{5}{c}{\textbf{Multitask Learning (Block + Frame-Level Shuffle)}} \\
        \hline
        \textbf{JiTTER (Multitask) - Best} & 0.75 & 0.5  & 0.25 & \textbf{0.574} \\
        JiTTER (Multitask) & 0.5  & 0.5  & 0.25 & 0.570 \\
        JiTTER (Multitask) & 0.75 & 0.25 & 0.25 & 0.567 \\
        JiTTER (Multitask) & 0.75 & 0.75 & 0.25 & 0.571 \\
        JiTTER (Multitask) & 0.75 & 0.5  & 0.5  & 0.573 \\
        \hline
    \end{tabular}
\end{table}

To evaluate the effectiveness of the proposed JiTTER framework, we compare it with the baseline MAT-SED model and conduct an ablation study on block-level shuffle, frame-level shuffle, and multitask learning. Additionally, we analyze the impact of block flipping and noise injection, then compare the best JiTTER with MAT-SED under controlled experimental conditions.

\subsection{Block-Level Shuffle}

Block-level shuffle aims to disrupt global temporal dependencies by perturbing the order of event segments while preserving all acoustic information. Unlike masked block prediction, which removes entire segments and requires the model to hallucinate missing content, block shuffle enforces explicit temporal reasoning by requiring the model to recover the correct event sequence rather than merely interpolating gaps. To evaluate its impact, we vary the block shuffle rate \( p_b \) while keeping frame-level shuffle disabled (\( p_{fb} = 0 \), \( p_{ff} = 0 \)). Each audio sequence consists of 100 time frames, which are partitioned into 20 non-overlapping blocks of size 5. The results are presented in Table \ref{tab:ablation}.

Block shuffle consistently improves PSDS over the baseline, with the highest improvement of 4.97\% at \( p_b = 0.75 \). This suggests that reconstructing shuffled event blocks forces the model to capture long-range temporal dependencies, which play a crucial role in transformer pretraining. The performance difference across different shuffle rates is relatively small, indicating that block shuffling provides stable improvements. However, excessive shuffling (\( p_b = 1.0 \)) leads to performance degradation, likely due to the introduction of excessive temporal disorder, making event reconstruction overly difficult.

One limitation of masked block prediction in MAT-SED is that it removes entire blocks, forcing the model to hallucinate missing content based on surrounding context \cite{matsed}. This can lead to unintended interpolation artifacts, where the model learns to reconstruct missing regions by exploiting statistical correlations rather than truly modeling event sequences. In contrast, JiTTER’s block shuffle retains all acoustic information while disrupting event ordering, requiring the model to reason explicitly about temporal coherence. This explicit enforcement of sequence reconstruction provides a more effective pretraining signal for SED.

\subsection{Frame-Level Shuffle}

Frame-level shuffle introduces fine-grained temporal perturbations, allowing the model to learn local transient dependencies. Unlike block-level shuffle, which primarily alters the event sequence at a coarse level, frame shuffle operates within individual event segments, perturbing intra-event structures while preserving the overall event order. This enables the model to refine event localization and better capture transient event characteristics. To analyze its effect, we vary the frame block selection rate (\( p_{fb} \)) and the fraction of shuffled frames per block (\( p_{ff} \)), while keeping block-level shuffle disabled (\( p_b = 0 \)). Each audio sequence consists of 100 time frames, which are divided into 5 non-overlapping blocks of size 20. The results are summarized in Table \ref{tab:ablation}.

Frame shuffle improves performance over the baseline, though its impact is less significant compared to block shuffling. The best configuration (\( p_{fb} = 0.5, p_{ff} = 0.25 \)) leads to a 3.8\% improvement. This suggests that while local temporal order is important, it has a smaller effect than global event structure. The results indicate that SED models benefit more from capturing event boundary structures at a coarser granularity (block level) rather than relying solely on fine-grained perturbations (frame level). Nevertheless, frame shuffle enhances event localization by introducing intra-block variability, which aids in transient event detection.

Unlike block shuffle, which restructures event sequences, frame shuffle focuses on intra-event variability by perturbing local structures without affecting the sequence at a higher level. This distinction suggests that block shuffle plays a more dominant role in shaping global event representations, whereas frame shuffle refines detailed event characteristics.

\subsection{Multitask Learning of Block and Frame-Level Shuffle}

We evaluate multitask learning, where block and frame-level shuffle are applied in separate training iterations. The results are presented in Table \ref{tab:ablation}. We experimented with the combination of the best block shuffle setting and the best frame setting, along with additional variations.

Multitask learning yields the highest PSDS of 0.574, improving by 5.71\% over the baseline. This suggests that combining global (block-level) and local (frame-level) perturbations enables more effective pretraining, as the model learns both event-level continuity and finer transient variations. Furthermore, the best-performing model arises not from simply stacking perturbations, but from a combination of optimal block0lvel and frame-level configurations. These findings highlight the importance of designing task-specific pretraining objectives rather than arbitrarily applying multiple augmentations.

Rather than relying solely on contextual interpolation, as seen in MAT-SED’s masked block prediction, JiTTER explicitly enforces temporal order reconstruction by reconstructing shuffled sequences. This prevents the model from over-relying on surrounding context and instead optimizes it to capture both fine-grained temporal order and global event structures. The results demonstrate that integrating perturbations at multiple temporal scales improves the model's ability to generalize across various sound event patterns. This highlights the advantage of explicitly modeling hierarchical temporal dependencies in SED pretraining.

\subsection{Block Flip in Block-level Shuffle}

\begin{table}[t]
    \centering
    \caption{Effect of block flip augmentation in JiTTER multitask learning.}
    \label{tab:blockflip}
    \begin{tabular}{l | c | c}
        \hline
        \textbf{Method} & flip rate & \textbf{PSDS} (\(\uparrow\)) \\
        \hline
        JiTTER (Multitask)      & -     & 0.574 \\
        \hline
        JiTTER (Multitask + Flip) & 0.25  & 0.572 \\
        JiTTER (Multitask + Flip) & 0.5   & 0.570 \\
        JiTTER (Multitask + Flip) & 0.75  & 0.571 \\
        \hline
    \end{tabular}
\end{table}

To further assess the impact of temporal transformations, we apply block flipping to the best multitask configuration. In this augmentation, shuffled blocks are reversed along the time axis with a probability defined by the hyperparameter \textit{flip rate}. The results are presented in Table \ref{tab:blockflip}. Contrary to expectations, block flipping does not improve performance and leads to a slight drop in PSDS. This suggests that excessive disruption of transient structures reduces the model’s ability to capture global temporal dependencies. One possible explanation is that block flipping distorts natural event progression by reversing localized patterns within audio sequences. 

From an auditory perception perspective, humans rarely encounter temporally inverted sound patterns in real-world scenarios. As a result, models trained with block flipping may learn non-representative patterns that do not generalize well to natural sound event structures. Additionally, phase incoherence introduced by time-reversed blocks may disrupt the model's ability to capture spectral-temporal relationships, which are crucial for precise event boundary detection.

Moreover, the pretext task of temporal reconstruction becomes significantly more challenging when flipped blocks are introduced. Unlike shuffled blocks, where the original order can be inferred through contextual reasoning, flipped blocks fundamentally alter the spectral envelope and transient structures of events, making it difficult for the model to learn meaningful representations. Similar findings have been observed in prior works on time-reversed signal processing, where artificially reversing signals degraded classification performance in audio and speech tasks.

\subsection{Noise Injection in Block-level Shuffle}

To explore a softer form of perturbation compared to block flipping, we investigate the impact of injecting Gaussian noise into shuffled blocks. Unlike masked prediction tasks, which remove entire regions of input audio, noise injection slightly obscures information while preserving weak structural cues. This allows the model to enhance feature robustness without fully disrupting temporal reconstruction.

\begin{table}[t]
    \centering
    \caption{Effect of noise injection on JiTTER multitask learning.}
    \label{tab:noise_injection}
    \begin{tabular}{l | c | c}
        \hline
        \textbf{Method} & Noise Scale \( \lambda \) & \textbf{PSDS} (\(\uparrow\)) \\
        \hline
        JiTTER (Multitask)      & -     & 0.574 \\
        \hline
        JiTTER (Multitask + Noise) & 0.05 & 0.570 \\
        JiTTER (Multitask + Noise) & 0.1  & \textbf{0.575} \\
        JiTTER (Multitask + Noise) & 0.2  & 0.567 \\
        JiTTER (Multitask + Noise) & 0.4  & 0.569 \\
        \hline
    \end{tabular}
\end{table}

We apply Gaussian noise sampled from \( \mathcal{N}(0, I) \) to shuffled blocks, scaling its magnitude by \( \lambda \). As shown in Table \ref{tab:noise_injection}, introducing mild noise (\( \lambda = 0.1 \)) improves PSDS to 0.575, surpassing the baseline multitask configuration (0.574). However, increasing \( \lambda \) beyond this threshold leads to performance degradation, with PSDS dropping to 0.567 at \( \lambda = 0.2 \) and 0.569 at \( \lambda = 0.4 \). These results suggest that a small degree of information corruption can regularize training and encourage better generalization. When applied in moderation, noise injection acts as a form of stochastic feature smoothing, preventing the model from overfitting to minute acoustic details that may not generalize well. 

This finding aligns with prior work in robust speech recognition and audio classification, where slight perturbations in the feature space have been shown to enhance generalization by encouraging invariance to minor spectral variations. The theoretical basis for this effect lies in stochastic regularization, where controlled noise injection forces the model to rely on robust acoustic features rather than overfitting to specific waveform characteristics. 

However, excessive noise injection disrupts key acoustic patterns, making it harder for the model to reconstruct meaningful temporal structures. As the noise magnitude increases, the signal-to-noise ratio (SNR) decreases, and event boundaries become harder to distinguish, leading to a decline in detection accuracy. This phenomenon aligns with perspectives in information theory, where too much noise obscures discriminative features necessary for accurate classification, ultimately degrading performance.

The results from both experiments indicate that block flipping and noise injection have fundamentally different effects on JiTTER’s pretraining dynamics. Block flipping disrupts the fundamental auditory structure by inverting event sequences, making it difficult for the model to reconstruct coherent time-aligned representations. This forces the model to develop sequence reordering capabilities but can also introduce unnatural phase distortions that degrade generalization. Noise injection, in contrast, preserves event order while slightly perturbing feature representations. This acts as a form of robustness enhancement rather than a restructuring task, promoting generalization without forcing the model to learn unrealistic event sequences.

These findings highlight an important consideration for designing self-supervised pretraining objectives: perturbations should balance informative learning signals with the risk of introducing non-naturalistic distortions. Our study suggests that structured shuffling, such as JiTTER’s hierarchical shuffle strategy, is a more effective SSL task than aggressive augmentations like full inversion or high-magnitude noise injection.

While noise injection provides a useful regularization mechanism, its impact remains secondary to structured temporal perturbations like block and frame-level shuffle. Future research could explore adaptive noise scheduling strategies or context-aware noise injection to enhance its benefits while minimizing potential degradation.

\subsection{Comparison with MAT-SED with Maximum PSDS}

\begin{table}[t]
    \centering
    \caption{Comparison of maximum PSDS scores across training runs.}
    \label{tab:max_psds}
    \begin{tabular}{l | c}
        \hline
        \textbf{Method} & \textbf{Max PSDS} (\(\uparrow\)) \\
        \hline
        MAT-SED \cite{matsed} & 0.587 \\
        MAT-SED (Reproduced) & 0.552 \\
        JiTTER (Best) & 0.584 \\
        \hline
    \end{tabular}
\end{table}

To ensure a fair comparison, we report the maximum PSDS scores measured across multiple training runs in Table~\ref{tab:max_psds}. While JiTTER achieves a maximum PSDS of 0.584, slightly below the originally reported MAT-SED result (0.587), it outperforms our reproduced MAT-SED baseline (0.552), demonstrating the robustness of our approach. The discrepancy between our reproduced MAT-SED score and the originally reported result can be attributed to three main factors:

\begin{enumerate}
    \itemsep 0.1em
    \item Differences in training infrastructure, such as GPU models, software versions, and CUDA configurations.
    \item Heuristically optimized hyperparameters may not generalize consistently across different environments.
    \item Random variations in the generation of the synthetic strongly labeled DESED dataset used for training.
\end{enumerate}

These factors underscore the challenges of reproducing results in deep learning and emphasize the need for fair comparisons under controlled settings.

\section{Conclusion}
\label{sec:conclusion}
In this work, we introduced JiTTER (Jigsaw Temporal Transformer for Event Reconstruction), a self-supervised learning framework designed to enhance temporal representation learning for SED. Unlike masked block prediction, which removes entire segments and forces interpolation, JiTTER preserves all information while enforcing explicit event reordering through hierarchical temporal shuffle reconstruction, enabling the model to capture both global event structures and fine-grained transient details. Experiments on the DESED dataset show that JiTTER achieves a 5.89\% PSDS improvement over MAT-SED, with ablation studies confirming that block-level shuffle strengthens long-range dependencies, frame-level shuffle enhances transient event detection, and multitask learning combining both perturbations yields the highest gains. Additional investigations reveal that block flipping disrupts essential event ordering, degrading performance, while moderate noise injection acts as a useful regularization mechanism. These findings highlight the importance of structured pretraining objectives that maintain event integrity while enforcing temporal reasoning. Beyond SED, JiTTER's approach is applicable to broader auditory tasks such as audio captioning, speaker recognition, and bioacoustic monitoring. Our results demonstrate that temporal order reconstruction is a crucial pretraining signal for event-aware SSL, providing a stronger alternative to conventional masking strategies.

\vspace{11pt}

\bibliographystyle{IEEEtran}
\bibliography{refs}

\begin{IEEEbiography}[{\includegraphics[width=1in,height=1.25in,clip,keepaspectratio]{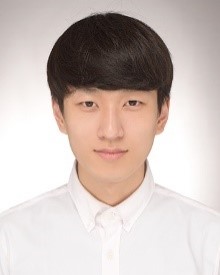}}]{Hyeonuk Nam}
received B.S. and M.S. degrees in mechanical engineering from Korea Advanced Institute of Science and Technology, Daejeon, Korea, in 2018 and 2020 respectively. He is currently pursuing the Ph.D. degree in mechanical engineering at the same institute.
His research interest includes various auditory intelligence themes including sound event detection, sound event localization and detection, automatic audio captioning, sound scene synthesis and human auditory perception.
\end{IEEEbiography}

\vspace{11pt}

\begin{IEEEbiography}[{\includegraphics[width=1in,height=1.25in,clip,keepaspectratio]{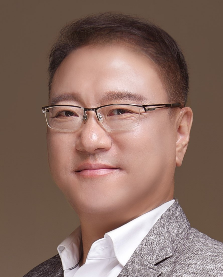}}]{Yong-Hwa Park}
received BS, MS, and PhD in Mechanical Engineering from KAIST in 1991, 1993, and 1999, respectively. In 2000, he joined to Aerospace Department at the University of Colorado at Boulder as a research associate. From 2003-2016, he worked for Samsung Electronics in the Visual Display Division and Samsung Advanced Institute of Technology (SAIT) as a Research Master in the field of micro-optical systems with applications to imaging and display systems. From 2016, he joined KAIST as professor of NOVIC+ (Noise \& Vibration Control Plus) at the Department of Mechanical Engineering devoting to research on vibration, acoustics, vision sensors, and condition monitoring with AI.
His research fields include structural vibration; condition monitoring from sound and vibration using AI; health monitoring sensors; and 3D sensors, and lidar for vehicles and robots. He is the conference chair of MOEMS and miniaturized systems in SPIE Photonics West since 2013. He is a vice-president of KSME, KSNVE, KSPE, and member of IEEE and SPIE

\end{IEEEbiography}

\vfill

\end{document}